\documentclass[3p,times]{elsarticle}

\usepackage{ecrc}

\volume{00}

\firstpage{1}

\journalname{Nuclear Physics A}

\runauth{}

\jid{nupha}

\usepackage{amssymb}

\usepackage{lineno}

\usepackage[figuresright]{rotating}

\begin{document}
\newcommand{\Npart}{\mbox{$\langle N_{\mathrm{part}}\rangle$}}
\newcommand{\Ncoll}{\mbox{$\langle N_{\mathrm{coll}}\rangle$}}
\newcommand{\dndeta}{\mbox{$dN_{ch}/d\eta$}}
\newcommand{\energy}{\mbox{$\sqrt{s_{NN}}=2.76$~TeV}}
\newcommand{\hienergy}{\mbox{$\sqrt{s_{NN}}=7$~TeV}}
\newcommand{\sqn}{\mbox{$\sqrt{s_{NN}}$}}
\newcommand{\sqs}{\mbox{$\sqrt{s}$}}
\newcommand{\pp}{\mbox{p+p}}
\newcommand{\sumet}{\mbox{$\sum E_T$}}
\newcommand{\rcp}{\mbox{$R_{\mathrm{CP}}$}}
\newcommand{\Cent}{\mbox{$\mathcal{C}$}}
\newcommand{\pT}{\mbox{$p_{\mathrm{T}}$}}
\newcommand{\pt}{\mbox{$p_{\mathrm{T}}$}}
\newcommand{\et}{\mbox{$E_{\mathrm{T}}$}}
\newcommand{\ptr}{\mbox{$p_{T}^{jet}/p_{T}^{Z}$}}
\newcommand{\zee}{\mbox{$Z\rightarrow e^{+}e^{-}$}}
\newcommand{\Zee}{\mbox{$Z\rightarrow e^{+}e^{-}$}}
\newcommand{\zmm}{\mbox{$Z\rightarrow\mu^{+}\mu^{-}$}}
\newcommand{\Zmm}{\mbox{$Z\rightarrow\mu^{+}\mu^{-}$}}
\newcommand{\ET}{\mbox{$E_T$}}
\newcommand{\massmm}{\mbox{$m_{\mu\mu}$}} 
\newcommand{\zll}{\mbox{$Z\rightarrow l^{+}l^{-}$}}
\newcommand{\ptZ}{\mbox{$p_{\mathrm{T}}^{Z}$}}
\newcommand{\ptjet}{\mbox{$p_{\mathrm{T}}^{\mathrm{jet}}$}}
\newcommand{\tnp}{\mbox{T\&P}}
\newcommand{\yZ}{\mbox{$y^{Z}$}}

\newcommand{\PbPb}{Pb+Pb}%
\newcommand{\inb}{\mbox{nb$^{-1}$}}
\def\TeV{\ifmmode {\mathrm{\ Te\kern -0.1em V}}\else
                   \textrm{Te\kern -0.1em V}\fi}%
\def\GeV{\ifmmode {\mathrm{\ Ge\kern -0.1em V}}\else
                   \textrm{Ge\kern -0.1em V}\fi}%

\begin{frontmatter}



\dochead{}

\title{Measurement of Z boson Production in Lead-Lead Collisions at \sqn=2.76,\TeV\ with the ATLAS Detector}


\author{Zvi Citron on behalf of the ATLAS Collaboration}

\address{}

\begin{abstract}
The ATLAS experiment has observed 1995 $Z$ boson candidates in 0.15\,\inb~ of integrated luminosity obtained in the 2011 LHC Pb+Pb~ run at \sqn=2.76 \TeV. The $Z$ bosons are reconstructed via di-electron and di-muon decay channels. The results from both channels are consistent with each other and are combined. The background is less than 3\% of the selected sample. Within the statistical and systematic uncertainties, the per-event $Z$ boson yield integrated over rapidity $|\yZ|<2.5$ is proportional to the number of binary collisions estimated by the Glauber model. The elliptic flow coefficient of the azimuthal distribution of the $Z$ boson with respect to the event plane is consistent with zero.
\end{abstract}


\end{frontmatter}


\section{Introduction}\label{sec:itro}
Extensive heavy ion (HI) programs carried out by the experiments at the Relativistic Heavy Ion Collider (RHIC) at BNL, and the Large Hadron Collider (LHC) at CERN, have established that the hot and dense matter produced in high energy collisions of HI imposes a significant energy loss on the energetic color charge carriers penetrating such a medium~\cite{RHIC_review,LHC_review}. An understanding of this phenomenon requires measuring the unmodified production rates of the particles in HI collisions before they lose energy. The best candidates to perform such measurements are particles which do not interact via the strong force. 

This note presents a measurement of $Z$ boson production in Pb+Pb collisions at \energy\ using the electron and muon decay channels. The results demonstrate that the $Z$ boson production rate in HI collisions is proportional to the mean number of binary nucleon-nucleon collisions (\Ncoll) calculated within the framework of the Glauber model~\cite{Miller:2007ri}. The analysis of the elliptic flow coefficient $v_{2}$ of the $Z$ boson angular distribution reveals no correlation between the emission angle of the particle and the azimuthal direction of the event plane.

The ATLAS detector~\cite{Aad:2008zzm} at the LHC covers nearly the entire solid angle around the collision point. It consists of an inner tracking detector surrounded by a thin superconducting solenoid, electromagnetic and hadronic calorimeters, and a muon spectrometer incorporating three superconducting toroid magnets. 

The inner-detector (ID) is immersed in a 2~T axial magnetic field and provides charged particle tracking in the pseudorapidity range $|\eta|<2.5$.
A high-granularity silicon pixel detector covers the vertex region and it is followed by a silicon microstrip tracker and a transition radiation tracker.

The calorimeter system covers the range $|\eta|< 4.9$. Within the region $|\eta|< 3.2$, electromagnetic calorimetry is provided by barrel and end-cap high-granularity lead liquid-argon (LAr) calorimeters, with an additional thin LAr presampler covering $|\eta| < 1.8$.  Forward calorimeters (FCal) are located in the range  $3.1<|\eta|<4.9$.

The muon spectrometer (MS) comprises separate trigger and precision tracking chambers measuring the deflection of muons in the magnetic field generated by superconducting air-core toroids. The precision chamber system covers the region $|\eta| < 2.7$ with three layers of monitored drift tubes~(MDT), complemented by cathode strip chambers~(CSC) in the innermost layer of the forward region, where the background is highest. The muon trigger system covers the range $|\eta| < 2.4$ with resistive plate chambers in the barrel, and thin gap chambers in the end-cap regions.

\section{Analysis}\label{sec:ana}
This analysis uses 2011 LHC lead-lead collision data collected with the ATLAS experiment and corresponding to an integrated luminosity of approximately 0.15\,\inb.

Electron candidates were identified at the first trigger level (L1) as a cluster formed with $(\Delta\phi\times\Delta\eta)=0.1\times0.1$ trigger towers of the electromagnetic calorimeter, covering the pseudorapidity range $|\eta|<2.5$, excluding the transition region between the barrel and end-cap calorimeters ($1.37<|\eta|<1.52$). Candidates were then identified using the standard ATLAS reconstruction algorithm~\cite{Aad:2011mk}, requiring the matching of a track to an electromagnetic energy cluster.    In addition to the track matching requirement,  cuts based on the balance between the track momentum and cluster energy ($E/p$) and shower shape variables, were also used for electron identification. Electron selection is limited to $|\eta|<2.5$ and both electrons are required to have $\et>20$\,\GeV.

Muon candidates were selected using all three levels of the trigger system. The L1 muon trigger searches for patterns of hits consistent with muons of a certain \pt\ in the trigger chambers within $|\eta|<2.4$. 
In addition, a full scan muon reconstruction was performed by the high level trigger to identify muons with $\pt>10$\,\GeV.  

Single muons were reconstructed with varying levels of quality~\cite{ATLAS-CONF-2010-036}. High quality muons were reconstructed in both the MS and ID subsystems with consistent angular measurements, as well as with a good match to the event vertex. At least one muon in each pair, matched to the trigger, is required to be of such quality. If the second muon in the pair has hit patterns in the MS and ID subsystems satisfying criteria of high reconstruction quality,  the minimum \pt\ threshold is set to 10\,\GeV\ on both muons. If the second muon fails this condition, both muons were required to satisfy $\pt>20$\,\GeV.

\begin{figure}[h!]
\begin{center}
\includegraphics[width=0.65\textwidth]{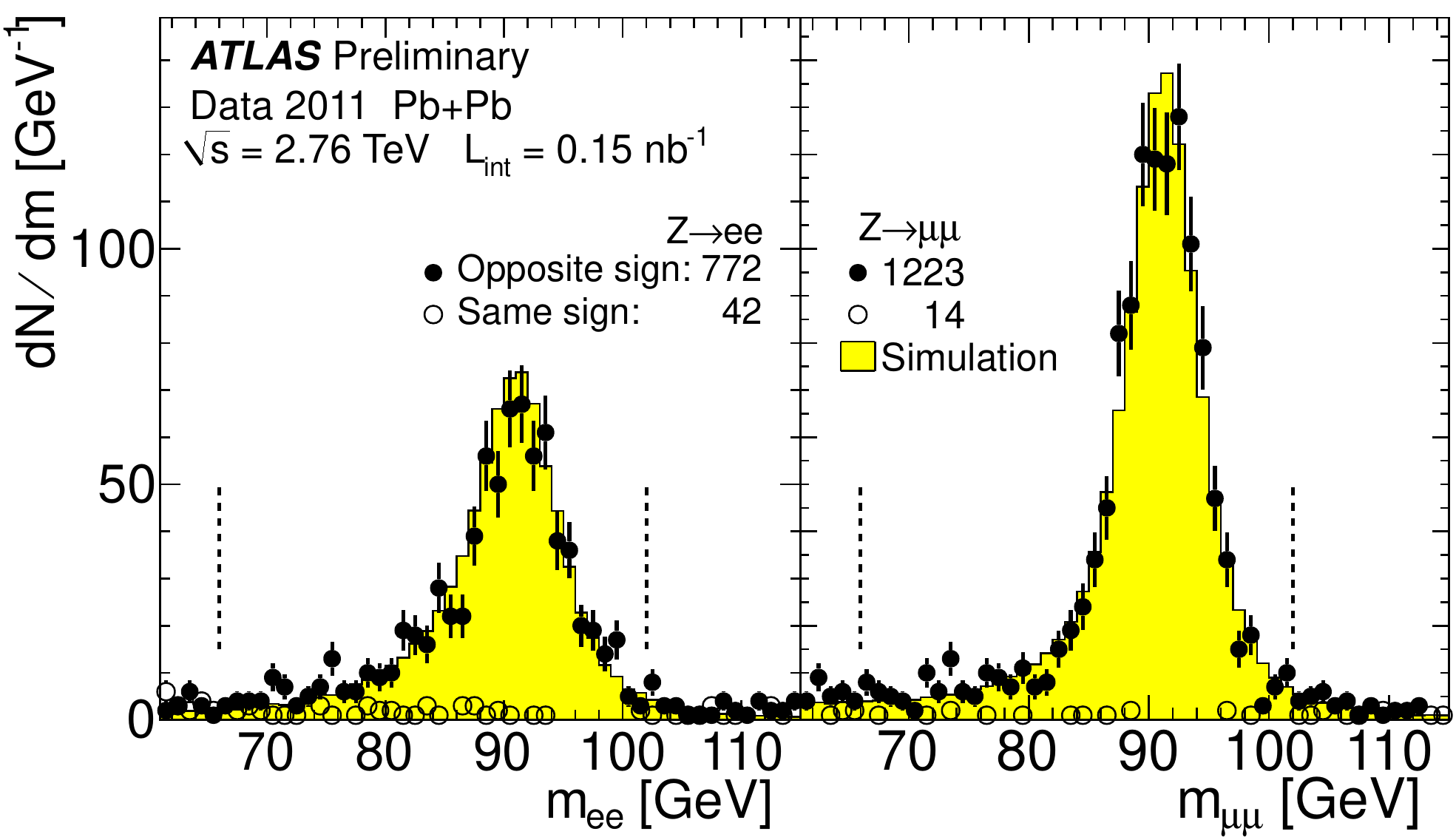}
\caption{The invariant mass distributions of \zee~ (left) and \zmm~ (right) in data (points) and simulation (histogram) integrated over momentum, rapidity, and centrality. The simulation is reweighted to match the centrality distribution in data and normalized in the region $66\GeV<m_{ll}<102$\,\GeV~ ($l=e,\mu$) indicated by dashed lines. Numbers of counts given in the plot correspond to the same mass region.}
\label{fig:mass_peaks}
\end{center}
\end{figure}

Once identified, electron and muon candidates were then paired in ee or $\mu\mu$ opposite charge combinations.  Pairs with an invariant mass between 66 and 102 \GeV\ were accepted as $Z$ boson candidates.  The $Z$ boson sample has a contamination from background processes of approximately 5\% in the \zee\ channel and 1\% in the \zmm\ channel, mostly from combinatorial sources.
The invariant mass distributions of the selected events together with estimated combinatorial backgrounds are shown in Fig.~\ref{fig:mass_peaks}.
Backgrounds from electroweak processes and top pair decays~\cite{Aad:2011dm} are small compared to the combinatorial background, and their contribution is neglected.
The $Z$ reconstruction efficiency is based on study of {\sc Pythia}  (version 6.425) ~\cite{Sjostrand:2006za}\ \zll\ events  with $66 \GeV<m_{Z}<116$\,\GeV\ and $|\yZ|<2.5$ embedded into \PbPb\ events generated by the {\sc Hijing} (version 1.38b) ~\cite{PhysRevD-44-3501} event generator. The response of the ATLAS detector to the generated particles was modeled using GEANT4~\cite{Agostinelli2003250,ATLASSim}. 

The main sources of systematic uncertainties on the measured yields in both lepton channels are associated with the precision to which the corrections applied to the data can be calculated. The efficiency and resolution uncertainties depend on the $Z$ boson kinematics and are on average 8\% (5.5\%) and 2.5\% (2.5\%), respectively for \zee (\zmm).   

Another way to search for any effect on \zll\ production due to the medium is to study the elliptic flow, $v_2$\cite{v2_def}, of the $Z$ boson. The $v_2$ is defined as the amplitude of the second Fourier harmonic of the $Z$ boson azimuthal emission angle with respect to the event plane.  The event plane contains the momentum vectors of the colliding nuclei, and is measured on an event-by-event basis from the azimuthal distribution of energy deposition in the FCal detectors.

\section{Results}\label{sec:res} 
\begin{figure}[htb]
\begin{center}
\includegraphics[width=0.85\textwidth]{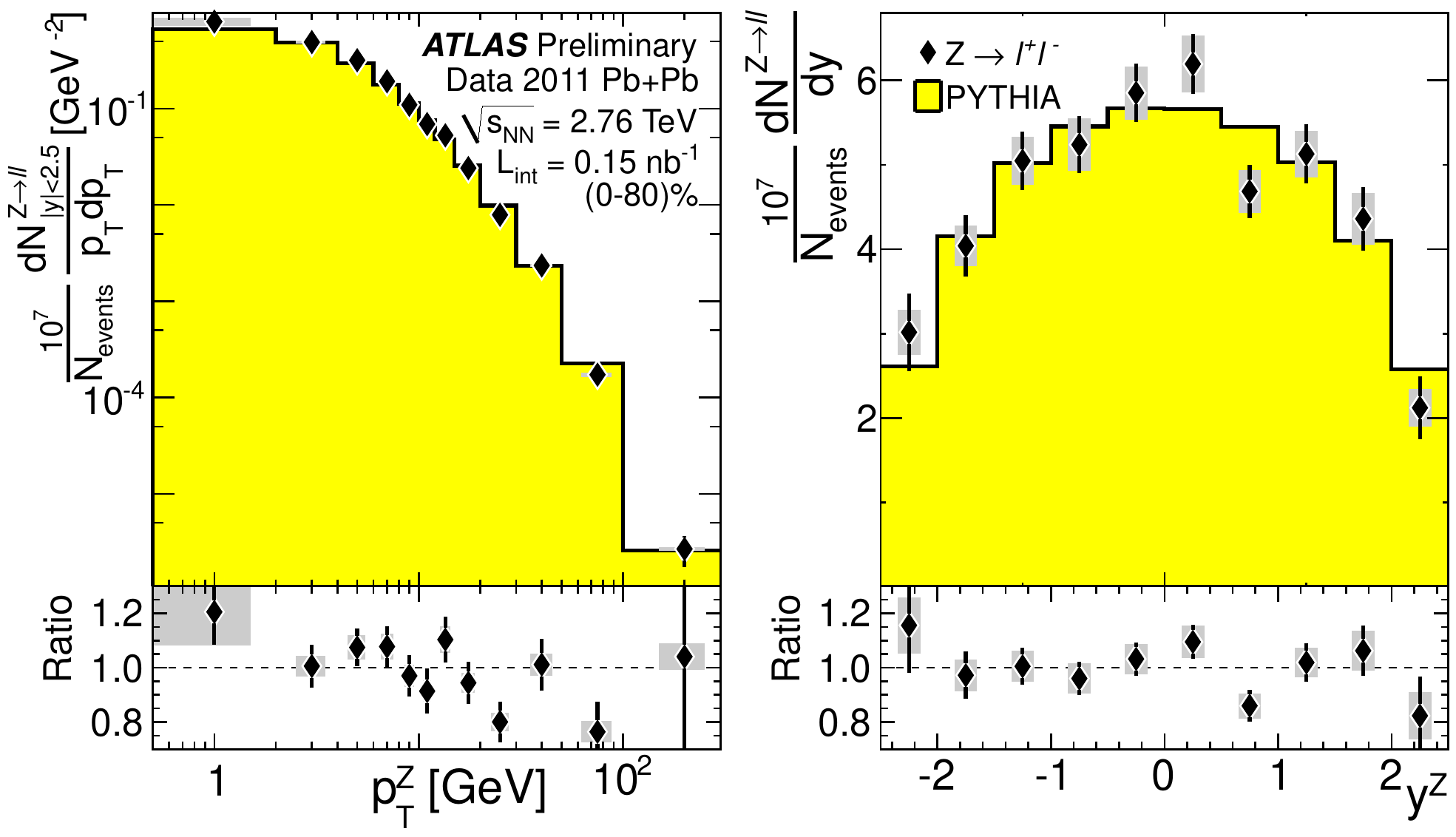}
\caption{Corrected transverse momentum (left) and rapidity (right) distributions of the measured $Z\rightarrow l^{+}l^{-}$ ($l=e,\mu$) data compared to {\sc Pythia} predictions normalized by area. The lower panels display the ratio of data over {\sc Pythia}.}
\label{fig:pt_rap}
\end{center}
\end{figure}

\begin{figure}[h!]
\begin{center}
\includegraphics[width=0.7\textwidth]{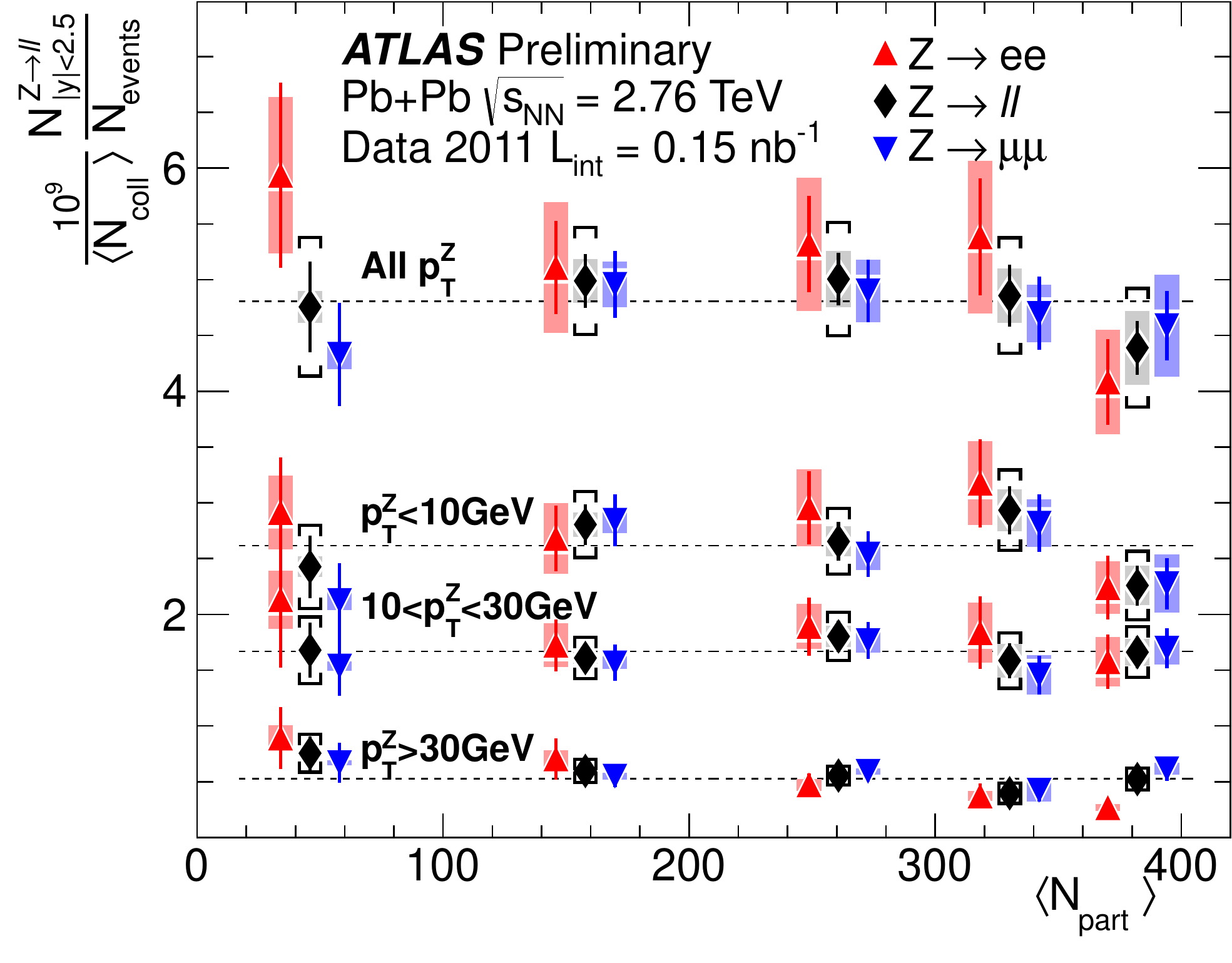}
\caption{Centrality dependence of $Z$ boson yields divided by \Ncoll, measured in $|\yZ|<2.5$. Results for $ee$  and $\mu\mu$ channels are shifted left and right respectively from their weighted average, which is plotted at the nominal \Npart~ value. The statistical (bars) and systematic (shaded bands) uncertainties are calculated using the appropriately weighted average of the two contributing sources. Brackets show the combined uncertainty including the uncertainty on \Ncoll. The dashed lines are constant fits to the combined results.} 
\label{fig:scaling}
\end{center}
\end{figure}

For both \zee~ and \zmm~ analyses, correction factors to account for the efficiency and detector resolution within the selected acceptance based on the simulation are calculated differentially in event centrality, \ptZ, and \yZ.  In each decay channel, the correction factor is applied and the corrected background, estimated by the same-sign pairs, is subtracted.  The two decay channel measurements are then combined with weights set by their respective uncertainties. The fully corrected \ptZ~ and \yZ~ distributions for the $Z$ boson are shown in Fig.~\ref{fig:pt_rap}. The data distributions plotted in (0-80)\% centrality agree in shape with the {\sc Pythia} simulations of $Z$ boson production in $pp$ collisions.\\
The $Z$ boson yields per-event, divided by \Ncoll, are shown in Fig.~\ref{fig:scaling}. 
The figure demonstrates that the \zee~ and \zmm~ results are consistent as a function of \Npart\ within their uncertainties for all \ptZ~ and centrality regions.  A combined uncertainty, including common uncertainties on the determination of \Ncoll~ is presented with the horizontal brackets. Within the statistical significance of the data sample, the $Z$ boson per-event yield obeys binary scaling. The dashed lines represent constant fits to the combined data as a function of the mean number of nucleons participating in the event, \Npart.

The $v_2$ of the $Z$ boson in the (0-60)\% centrality interval is measured to be $-0.015 \pm 0.018$(stat.)$\pm 0.014$(sys.).  This observation is an independent measurement consistent with \zll\ yields being unaffected by the medium in HI collisions.

\section{Conclusions}\label{sec:con}
$Z$ boson production has been measured in \PbPb~ collisions at \sqn=2.76\,\TeV~ using 0.15\,\inb~ of integrated luminosity collected in the 2011 LHC run. Within $|\yZ|<2.5$, and $66\GeV<m_{ll}<102$\,\GeV, a total of 772 and 1223 $Z$ boson candidates are reconstructed in the \zee~ and \zmm~ channels respectively. The combinatorial background is at the level of 5\% in the  electron channel and 1\% for the muon channel. Yields of the $Z$ boson production integrated within $|y^{Z}|<2.5$ are consistent in the two channels in all measured \pt~ and centralities. The momentum and rapidity distributions of the $Z$ bosons measured in (0-80)\% centrality region are consistent in shape with the {\sc Pythia} simulations of $Z$ boson production in $pp$ collisions. Within the uncertainties the $Z$ boson yield is found to be proportional to \Ncoll. The elliptic flow coefficient  of the $Z$ boson is consistent with zero within the uncertainties of the measurements. 






\end{document}